\documentclass[preprint,aps,prd,groupedaddress,showpacs,nofootinbib]{revtex4-1}%
\usepackage{graphicx}
\usepackage{amsmath}
\usepackage{amssymb}
\usepackage{bm}
\usepackage{epsfig}
\usepackage{dcolumn}
\usepackage{slashed}
\usepackage{color}

\def\NLO{\text{NLO}}
\def\NNLO{\text{NNLO}}

\begin{document}

\def\Maryland{Maryland Center for Fundamental Physics, University of Maryland, College Park, Maryland 20742, USA}
\def\Argonne{High Energy Physics Division, Argonne National Laboratory, Argonne, IL 60439, USA}
\def\Northwestern{Department of Physics \& Astronomy, Northwestern University, Evanston, IL 60208, USA}

\title{A comparison of NNLO QCD predictions with 7 TeV ATLAS and CMS data for $V$+jet processes}

\author{Radja Boughezal}
\email{rboughezal@anl.gov}
\affiliation{\Argonne}

\author{Xiaohui Liu}
\email{xhliu@umd.edu}
\affiliation{\Maryland}

\author{Frank Petriello}
\email{f-petriello@northwestern.edu}
\affiliation{\Argonne}
\affiliation{\Northwestern}

\begin{abstract}
We perform a detailed comparison of next-to-next-to-leading order (NNLO) QCD predictions for the $W$+jet and $Z$+jet processes with 7 TeV experimental data from ATLAS and CMS.  We observe excellent agreement between theory and data for most studied observables, which span several orders of magnitude in both cross section and energy.  For some observables, such as the $H_T$ distribution, the NNLO QCD corrections are essential for resolving existing discrepancies between theory and data.

\end{abstract}

\maketitle

\section{Introduction}

The production of a vector boson in association with jets plays a critical role in the physics program of the Large Hadron Collider (LHC).  These processes can be measured with small errors over a large energy range, and have the potential to provide stringent tests of the Standard Model.  Significant theoretical effort has been devoted to understanding these processes.  During Run I of the LHC the ATLAS and CMS collaborations probed jet momenta in both $W$+jet and $Z$+jet reaching 1 TeV, and compared their 7 TeV results with a wide range of QCD predictions.  This brought the precision comparison of the Standard Model electroweak sector with data into a new energy range.  In addition to precision tests of the Standard Model, vector boson plus jet processes are backgrounds to some of the most important signatures of new physics at the LHC, including supersymmetry and dark matter.

The importance of and interest in vector boson plus jet production has long made it a target for detailed theoretical study, and there has been remarkable recent progress  in our ability to precisely predict both the $W$+jet and $Z$+jet processes.
The next-to-leading order (NLO) electroweak corrections were considered in Refs.~\cite{EWcorW,EWcorZ}, and  a merged NLO QCD+electroweak prediction was obtained in Ref.~\cite{NLOQCDEW}.  Very recently the full next-to-next-to-leading order (NNLO) QCD predictions have been derived~\cite{Boughezal:2015dva,Ridder:2015dxa,Boughezal:2015ded}.  These results make possible a new quantitative frontier in the comparison of LHC jet measurements with theory.  

It is our goal in this manuscript to perform a detailed comparison of NNLO theoretical predictions for vector boson plus jet production with 7 TeV measurements performed by the ATLAS and CMS collaborations.   ATLAS and CMS have compared their data with both NLO QCD predictions and several parton shower simulations that contain merged samples of leading-order $V$+n-jet amplitudes, or matching to a NLO $V$+jet calculation~\cite{Aad:2014qxa,Khachatryan:2014uva,Aad:2013ysa,Khachatryan:2014zya}.  In general these predictions agree well with the data.  However, there are several notable exceptions where significant discrepancies are observed.  NNLO QCD resolves these issues and leads to excellent agreement between theory and experiment over the entire measured energy range and for all observables.  In particular, the $H_T$ distributions in both $W$+jet and $Z$+jet production, and at both ATLAS and CMS, exhibit poor agreement with the theoretical predictions. The NLO QCD predictions undershoot the data in the high-$H_T$ region, as do other approaches such as LoopSim~\cite{Rubin:2010xp} and exclusive sums.  Several of the parton shower simulations overshoot the high-$H_T$ data. Only NNLO QCD is in excellent agreement with all available data for the $H_T$ distribution.  Indeed, one of the major conclusions of our paper is the remarkable power of fixed-order QCD in describing the entire suite of $V$+jet measurements, which span orders of magnitude in both energy and cross section.

Our paper is organized as follows.  We describe our calculational setup, including the selection cuts for the ATLAS and CMS data against which we compare, in Section~\ref{sec:setup}.  Comparisons of theory with $W$+jet data from both ATLAS and CMS data are performed in Section~\ref{sec:Wnum}.  Similar comparisons against $Z$+jet data from ATLAS and CMS are done in Section~\ref{sec:Znum}.  Finally, we summarize and conclude in Section~\ref{sec:conc}.

\section{Setup}
\label{sec:setup}

We discuss here our calculational setup for vector boson production in association with a jet through NNLO in perturbative QCD.  We study collisions at a 7 TeV LHC, for which both ATLAS and CMS data is publicly available.    Jets are defined using the anti-$k_t$ algorithm~\cite{Cacciari:2008gp}, with a different distance measure for ATLAS and CMS as detailed later in this section.  We use CT14 parton distribution functions (PDFs)~\cite{Dulat:2015mca} at the appropriate order in perturbation theory: LO PDFs together with a LO partonic cross section, NLO PDFs with a NLO partonic cross section, and NNLO with a NNLO partonic cross section.  As indicated later in the text, we have computed some results using other PDF sets as well.  We choose the central scale 
\begin{equation}
\mu_0 = \sqrt{M_{V}^2+\sum_i (p_T^{J_i})^2}
\end{equation}
for both the renormalization and factorization scales, where $M_{V}$ is the invariant mass of the vector boson and the sum $i$ runs over all reconstructed jets.  This dynamical scale correctly captures the characteristic energy throughout the entire kinematic range studied here.  To estimate the theoretical uncertainty we vary the renormalization and factorization scales independently in the range $\mu_0/2 \leq \mu_{R,F} \leq 2 \mu_0$, subject to the restriction
\begin{equation}
1/2 \leq \mu_R / \mu_F \leq 2.  
\end{equation}
All numerical results presented for $W$-boson production include both $W^{+}$ and $W^{-}$ contributions, and all $Z$-boson results include the contribution of off-shell photon exchange.

Our selection criteria match those used by ATLAS and CMS in their publicly available studies of $W$+jet and $Z$+jet production.  For the $W$-boson selection we implement the cuts shown in Table~\ref{tab:wcuts}\footnote{We note that CMS imposes cuts on the jet pseudorapidity while ATLAS imposes cuts on the jet rapidity; although we use the notation $\eta$ for both ATLAS and CMS, we use the appropriate variable when comparing with each experiment.}.  For the ATLAS selection we have adopted the ``combined" cuts, where the muon and electron channels are extrapolated to a common phase space region.  The selection criteria for the $Z$-boson are presented in Table~\ref{tab:zcuts}.  The separation parameter $\Delta R$ is defined as 
\begin{equation}
\Delta R_{xy} = \sqrt{(\eta_x - \eta_y)^2+(\phi_x-\phi_y)^2},
\label{eq:delr}
\end{equation}
where $\phi_x$ denotes the transverse-plane azimuthal angle of particle $x$.  We note that there is an additional requirement that any jet falling within the region $\Delta R_{Jl} < 0.5$ is removed from the analysis.

\begin{table}[htbp]
\begin{tabular}{|c|c|c|}
\hline
$W$-boson cuts & ATLAS~\cite{Aad:2014qxa} & CMS~\cite{Khachatryan:2014uva} \\
 \hline\hline
lepton $p_T$ & $p_T^l > 25$ GeV & $p_T^l > 25$ GeV \\
lepton $\eta$ & $|\eta^l| < 2.5$ & $|\eta^l| < 2.1$\\
missing $E_T$ & $E_T^{miss} > 25$ GeV & $-$ \\
transverse mass & $m_T>40$ GeV & $m_T>50$ GeV \\
\hline
jet $p_T$ & $p_T^{J} > 30$ GeV & $p_T^{J} > 30$ GeV\\
jet $\eta$ & $|\eta^J| < 4.4$ &  $|\eta^J| < 2.4$\\
anti-$k_T$ radius & $R=0.4$ & $R=0.5$\\
\hline
\end{tabular}
\caption{Selection criteria for the $W$-boson plus jet cross section, following ATLAS and CMS 7 TeV analyses.}
\label{tab:wcuts}
\end{table} 

In their public analysis the ATLAS collaboration provides details regarding various correction factors they apply to the theoretical predictions.  These corrections account for the effects of hadronization and the underlying event, and are needed to transform the fixed-order theoretical predictions from the parton level to the particle level.  They also provide the corrections that account for final-state QED radiation.  These factors are provided for the leading-jet transverse momentum and rapidity distributions, as well as for the fiducial cross sections.  We apply the non-perturbative corrections obtained using the ALPGEN+HERWIG AUET2 tune~\cite{Aad:2014qxa,Aad:2013ysa}, and the QED factors from the muon channel.  We have checked that using the QED corrections associated with the electron channel instead lead to a negligible difference in our results.     The combined correction shifts the theoretical predictions by a few percent at most in the low-$p_T$ region, and quickly becomes negligible at high transverse momentum.

\begin{table}[htbp]
\begin{tabular}{|c|c|c|}
\hline
$Z$-boson cuts & ATLAS~\cite{Aad:2013ysa} & CMS~\cite{Khachatryan:2014zya} \\
 \hline\hline
lepton $p_T$ & $p_T^l > 20$ GeV & $p_T^l > 20$ GeV \\
lepton $\eta$ & $|\eta^l| < 2.5$ & $|\eta^l| < 2.4$\\
\hline
lepton separation &  $\Delta R_{ll} > 0.2$ & $-$ \\
lepton invariant mass & $66 \, \text{GeV} < m_{ll} < 116 \, \text{GeV}$ & $71 \, \text{GeV} < m_{ll} < 111 \, \text{GeV}$\\
\hline
jet $p_T$ & $p_T^{J} > 30$ GeV & $p_T^{J} > 30$ GeV\\
jet $\eta$ & $|\eta^J| < 4.4$ &  $|\eta^J| < 2.4$\\
anti-$k_T$ radius & $R=0.4$ & $R=0.5$\\
\hline
\end{tabular}
\caption{Selection criteria for the $Z$-boson plus jet cross section, following ATLAS and CMS 7 TeV analyses.}
\label{tab:zcuts}
\end{table} 

In our analysis we compare the theory predictions against data in the inclusive one-jet bin.  For the exclusive one-jet bin, it would be interesting to pursue a more detailed investigation that also considers the predictions of resummation-improved perturbation theory in addition to fixed-order NNLO QCD, such as performed in Ref.~\cite{Boughezal:2015oga}, and we postpone such a study to future work.  We compare the following distributions against the available data:  $p_T^{J_1}$, $p_T^Z$, $\eta^{J_1}$ and $H_T$ (we note that data for $p_T^W$ is not presented in the either experimental analysis).  Here, $\eta^{J_1}$ is the rapidity (for ATLAS) or pseudorapidity (for CMS) of the leading jet and $p_T^{J_1}$ is the transverse momentum of the leading jet.  $p_T^Z$ is the transverse momentum of the reconstructed dilepton system in the $Z$+jet channel.  $H_T$ is the scalar sum of the transverse momenta of all reconstructed jets\footnote{We note that this quantity is instead called $S_T$ by ATLAS.}.  All of these distributions begin first at leading order for the $V$+jet process, and therefore the results presented here are genuine NNLO predictions.  Other distributions that require two jets are only described at NLO by our calculation.   Comparisons of NLO QCD calculations with the data have already been performed by the experimental collaborations, and we do not reproduce such studies here.

Our focus in this work is on the impact of NNLO QCD corrections, but we comment here on other sources of theoretical uncertainty, most notably those arising from PDFs and electroweak corrections.  In Ref.~\cite{Boughezal:2016isb} the effect of changing the PDF set on $Z$+jet production in 13 TeV collisions was studied.  The differences between the CT14~\cite{Dulat:2015mca}, NNPDF3.0~\cite{Ball:2014uwa}, and MMHT~\cite{Harland-Lang:2014zoa} sets were much smaller than the scale uncertainty errors for the $p_T^Z$, $p_T^{J_1}$, and $H_T$ distributions.  The ABM12 set~\cite{Alekhin:2013nda} exhibited differences larger than the scale uncertainties in these distributions, indicating that these observables may eventually be used to improve our knowledge of the PDFs.  We expect that these conclusions will also hold for the 7 TeV collisions considered here.   An up-to-date study of the electroweak corrections to $V$+jet was performed in Ref.~\cite{NLOQCDEW}.  Their impact is very observable dependent.  They are small, in the couple-to-few percent range, for the regions of $H_T$ and $p_T^{J_1}$ considered here.  The electroweak corrections do not have a large impact on the $\eta_J^1$ distribution.  For the $p_T^Z$ distribution the electroweak corrections can be important, reaching the 10\% level near the upper boundary of the region considered here.   Although this is a significant effect, it is within the experimetnal uncertainties, which reach nearly 20\% in the high-$p_T^Z$ region.  In the future the electroweak corrections should be combined with the NNLO QCD results studied here.

The NNLO calculation upon which our phenomenological study is based was obtained using the $N$-jettiness subtraction scheme~\cite{Boughezal:2015dva,Gaunt:2015pea}.  This technique relies upon splitting the phase space for the real emission corrections according to the $N$-jettiness event shape variable, $\tau_N$~\cite{Stewart:2010tn}, and relies heavily upon the theoretical machinery of soft-collinear effective theory~\cite{scet}. For values of $N$-jettiness greater than some cut, $\tau_N > \tau_N^{cut}$, NLO calculations for $W$+2-jets and $Z$+2-jets are used.  Any existing NLO program can be used to obtain these results.  We use MCFM~\cite{Campbell:2010ff,Campbell:2015qma} in this study.  For the phase-space region $\tau_N < \tau_N^{cut}$, an all-orders resummation formula is used to obtain the contribution to the cross section~\cite{Stewart:2010tn,Stewart:2009yx,Gaunt:2014xga,Becher:2006qw,Boughezal:2015eha}.  An important check of the formalism is the independence of the full result from $\tau_N^{cut}$.  By now the application and validation of $N$-jettiness subtraction for one-jet processes has been discussed several times in the literature~\cite{Boughezal:2015dva,Boughezal:2015aha,Boughezal:2015ded}, and we do not review this topic here.  We note that we have computed each bin of the studied distributions for several $\tau_N^{cut}$ values, and have found independence of all results from $\tau_N^{cut}$ within numerical errors.

\section{Numerical results for 7 TeV $W$+jet production}
\label{sec:Wnum}

We begin by discussing $W$-boson production in association with a jet.  We first compare the fiducial cross sections measured by ATLAS and CMS with both NLO and NNLO QCD predictions in Table~\ref{tab:fidW}, assuming the cuts of Table~\ref{tab:wcuts}.  The NNLO QCD corrections shift the NLO fiducial cross section by $+3\%$ for CMS cuts and leave the result nearly unchanged for ATLAS cuts.  For both cases the NNLO predictions are in good agreement with the experimental measurements, within errors.  For the CMS cuts the NNLO QCD correction brings the prediction into slightly better agreement with the measured result.  In the ATLAS case both the NLO and NNLO cross sections are slightly below the measured value, but are within the still-large $1\sigma$ experimental error band.  The residual scale variation is greatly reduced by the inclusion of the NNLO corrections, decreasing from the $\pm 5\%$ level at NLO to the $\pm1\%$ level at NNLO.

\begin{table}[h]
\begin{tabular}{|c|c|c||c|}
\hline
$W$-boson & $\sigma_{\NLO}$ (pb) & $\sigma_{\NNLO}$ (pb) & experiment (pb)  \\
 \hline\hline
ATLAS cuts: & $482^{+31}_{-26}$ & $483^{+0}_{-5}$ & $493.8^{+0.5 \text{(stat)}+ 43\,\text{(sys)} 
	+ 9.7\,\text{(lumi)}}_{-0.5 \text{(stat)}- 43\,\text{(sys)} 
	- 9.7\,\text{(lumi)}}$\\
CMS cuts: & $467^{+29}_{-24}$ & $481^{+0}_{-5}$ & $479.8^{+18.3}_{-19.6}$  \\
\hline
\end{tabular}
\caption{Fiducial cross sections for the inclusive $W$+1-jet bin for both ATLAS and CMS cuts.  The scale-variation errors are shown for the NLO and NNLO cross sections.}
\label{tab:fidW}
\end{table} 

\begin{figure}[htbp]
    \includegraphics[width=.75\linewidth]{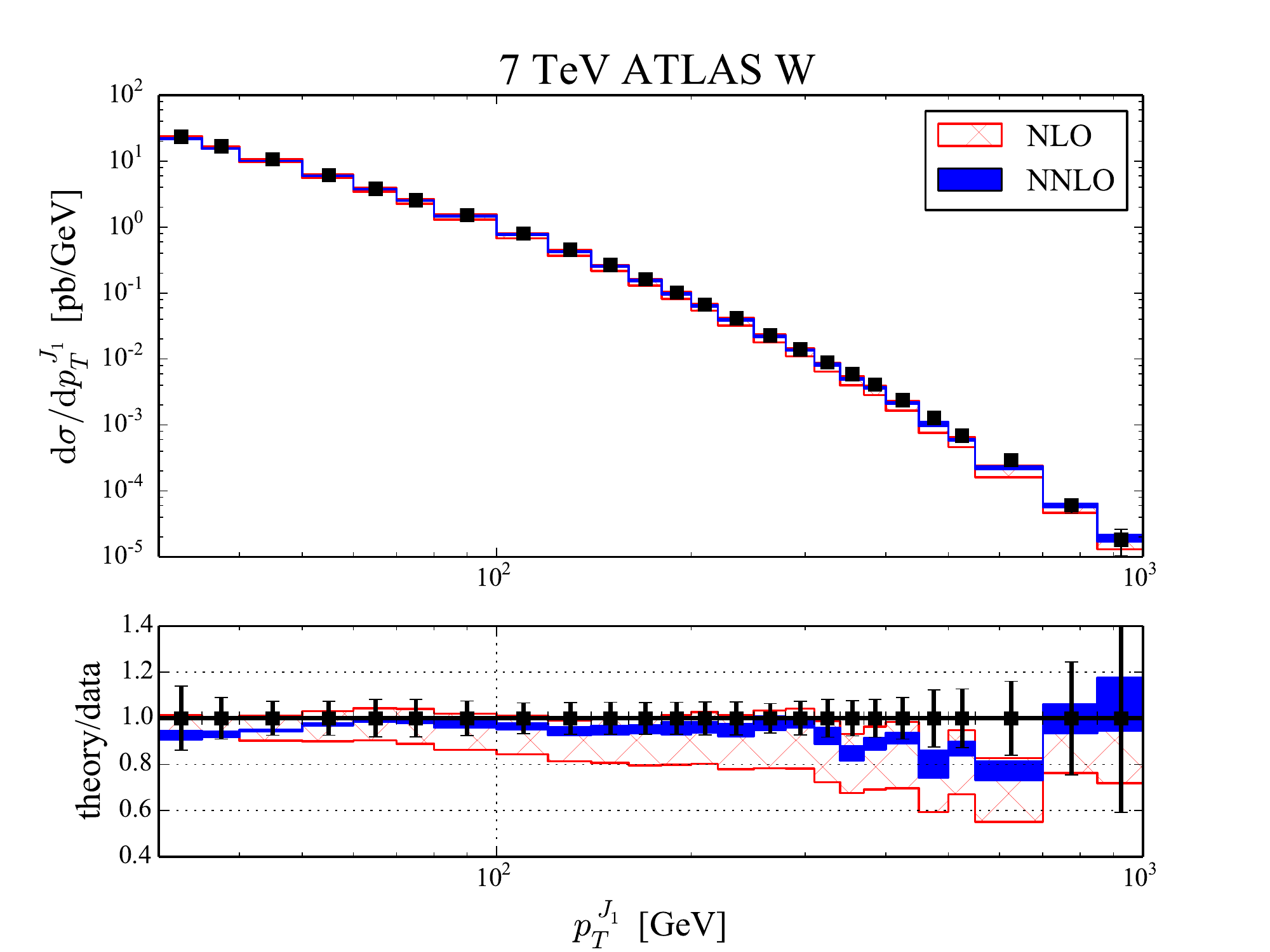} 
    \includegraphics[width=.75\linewidth]{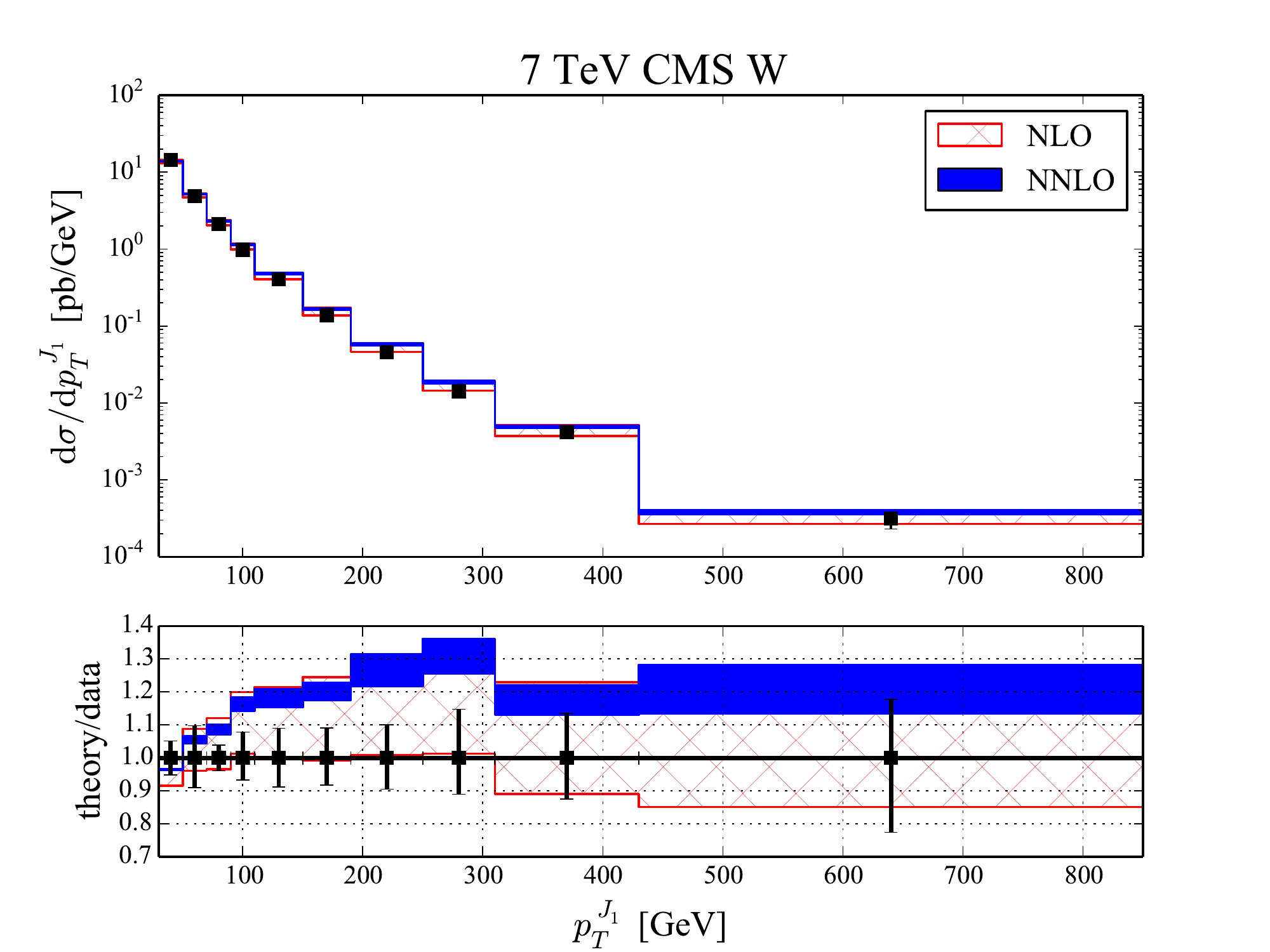} 
    \caption{Plots of the leading-jet transverse momentum distribution for ATLAS and CMS.  The upper panel of each plot shows the distributions at NLO and NNLO, as well as the experimental data points.  The lower panel of each plot shows the ratio of the NLO and NNLO predictions to the measured data.  The black error bars denote the experimental errors, the red hatched band denotes the NLO scale variation, and the blue solid band indicates the NNLO scale variation.}    
    \label{fig:pTJ1W}
\end{figure}

We next study the NLO and NNLO theoretical predictions for the transverse momentum distribution of the leading jet in Fig.~\ref{fig:pTJ1W}.  Our binning for this observable and for all other distributions follows exactly the binning used by the experimental collaborations.  We note that the wiggles seen in the lower panels for these plots, and in all other plots, arise from the errors in the experimental data, and not from the theoretical predictions.  Theory is in good agreement with the ATLAS data.  The theory slightly undershoots the data, similar to the behavior seen for the fiducial cross section in Table~\ref{tab:fidW}.  At intermediate to high transverse momenta the NNLO QCD corrections increase the prediction, leading to a better agreement with ATLAS data.  We note that ATLAS has compared their data against several theoretical predictions~\cite{Aad:2014qxa}: NLO QCD from Blackhat+Sherpa~\cite{Berger:2009ep}, the LoopSim approximation~\cite{Rubin:2010xp}, various tree-level predictions with multi-leg merging combined with parton showers, and the {\tt MEPS@NLO} approach~\cite{Hamilton:2010wh,Hoche:2010kg}.  They find that the NLO QCD, LoopSim and {\tt MEPS@NLO} predictions are all lower than the experimental data. Those based on merged tree-level amplitudes combined with a parton shower are slightly higher than the measurements, but still within experimental errors. Both NNLO QCD and the merged tree-level samples provide a good description of the data over the entire $p_T^{J_1}$ range.  We note that the scale variation errors are smaller than the experimental errors throughout the entire studied range.

Both the NLO and NNLO predictions are systematically higher than the CMS data starting at $p_T^{J_1} \approx 100$ GeV.  At NLO the large scale dependence of the theoretical prediction masks this discrepancy, but it becomes clear at NNLO, when the theory errors are reduced.  Similar discrepancies between merged leading-order plus parton-shower predictions and the CMS $p_T^{J_1}$ data are observed by the collaboration~\cite{Khachatryan:2014uva}.  The electroweak corrections are expected to only slightly decrease the higher-order QCD result in this energy range~\cite{NLOQCDEW}, and are therefore unlikely to resolve the discrepancy.  

\begin{figure}[htbp]
    \includegraphics[width=.75\linewidth]{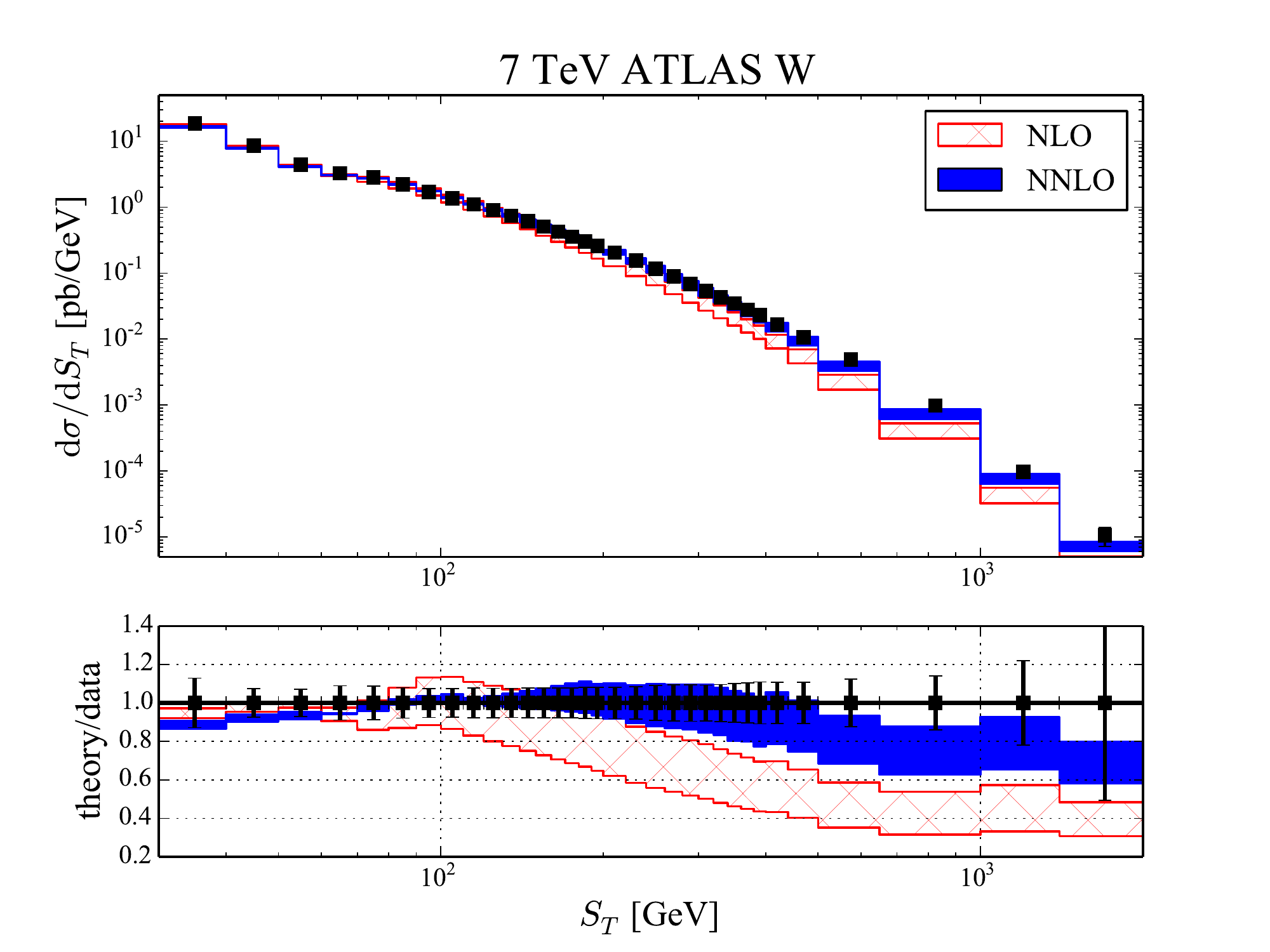} 
    \includegraphics[width=.75\linewidth]{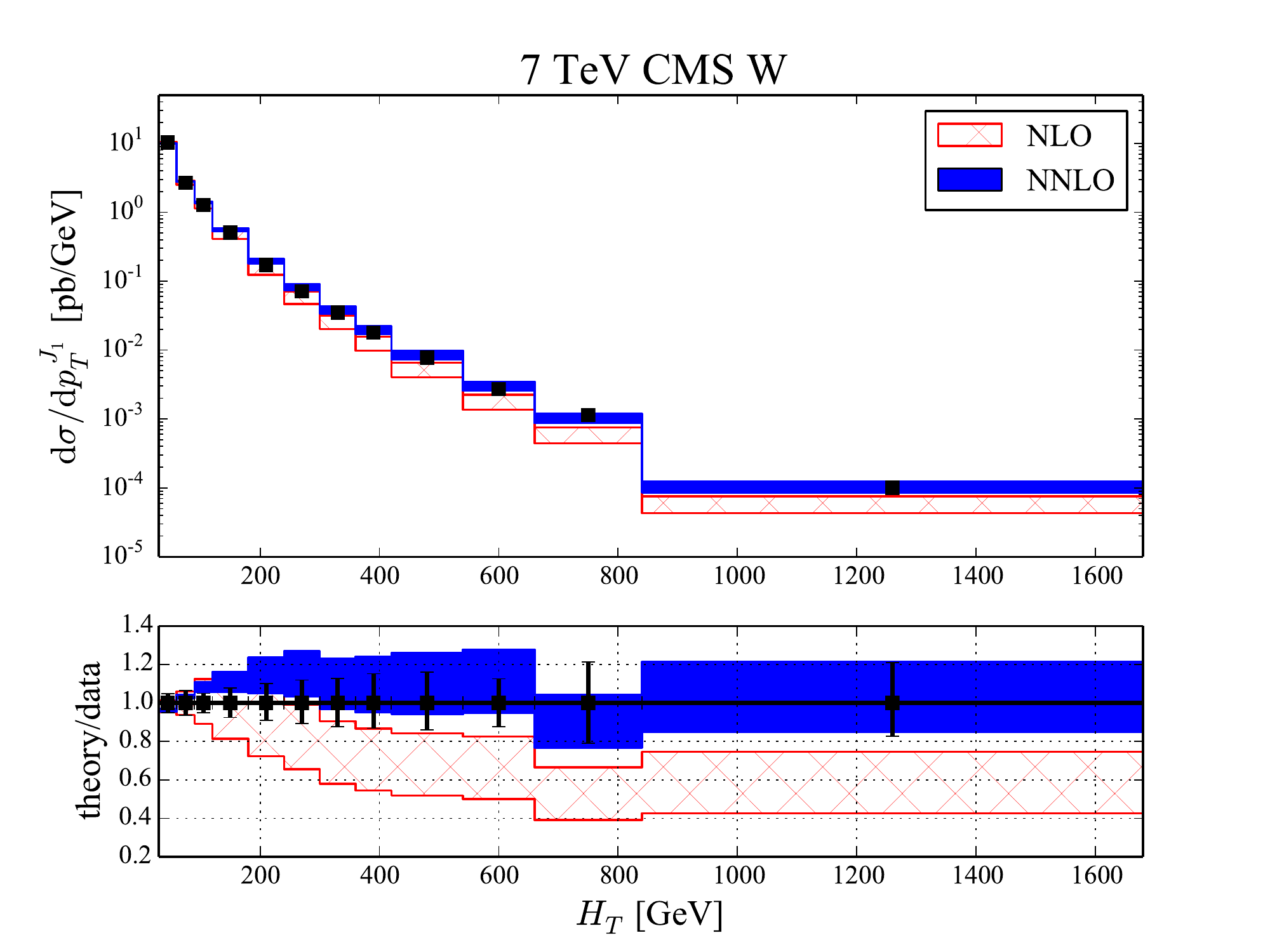} 
    \caption{Plots of the $H_T$ (CMS) and $S_T$ (ATLAS)  distributions, where both variables are defined as the scalar sum of jet transverse momenta.  The upper panel of each plot shows the distributions at NLO and NNLO, as well as the experimental data points.  The lower panel of each plot shows the ratio of the NLO and NNLO predictions to the measured data.  The black error bars denote the experimental errors, the red hatched band denotes the NLO scale variation, and the blue solid band indicates the NNLO scale variation.}
    \label{fig:HTW}
\end{figure}

In Fig.~\ref{fig:HTW} the comparison between theory and data for the $H_T$ distribution is shown\footnote{ATLAS calls this variable $S_T$, and we follow their notation in our plot label.}.  The need for NNLO QCD in the comparison to data is clear from these plots. The NLO predictions far undershoot the ATLAS and CMS predictions at high $H_T$.  The NNLO corrections bring theory into good agreement with experiment throughout the entire energy range, with only a slight undershoot at very high $H_T$ for the ATLAS comparison where the experimental errors are large.  The reason for these large corrections has been discussed in the literature~\cite{Rubin:2010xp}.  At NLO there are configurations containing two hard jets and a soft/collinear $W$-boson that are logarithmically enhanced. Such contributions cannot occur at LO, since the $W$-boson must balance in the transverse plane against the single jet that appears.   An accurate theoretical prediction for these configurations is first obtained upon inclusion of the NNLO corrections. Similarly, the combination of these dijet contributions with the events containing a high-$p_T$ $W$-boson in the correct proportion  is only obtained at NNLO in QCD perturbation theory.  Both ATLAS and CMS compare their data against the NLO QCD prediction from Blackhat+Sherpa and against several simulations containing fixed-order cross sections combined with a parton shower.  These predictions overshoot the data at high $H_T$, except for the {\tt MEPS@NLO} approach which undershoots the data at intermediate $H_T$.  Only NNLO QCD can reliably predict this distribution over the entire energy range.  We note that the estimated NNLO theoretical errors and the experimental errors are comparable in the intermediate and high $H_T$ regions.

\begin{figure}[htbp]
    \includegraphics[width=.75\linewidth]{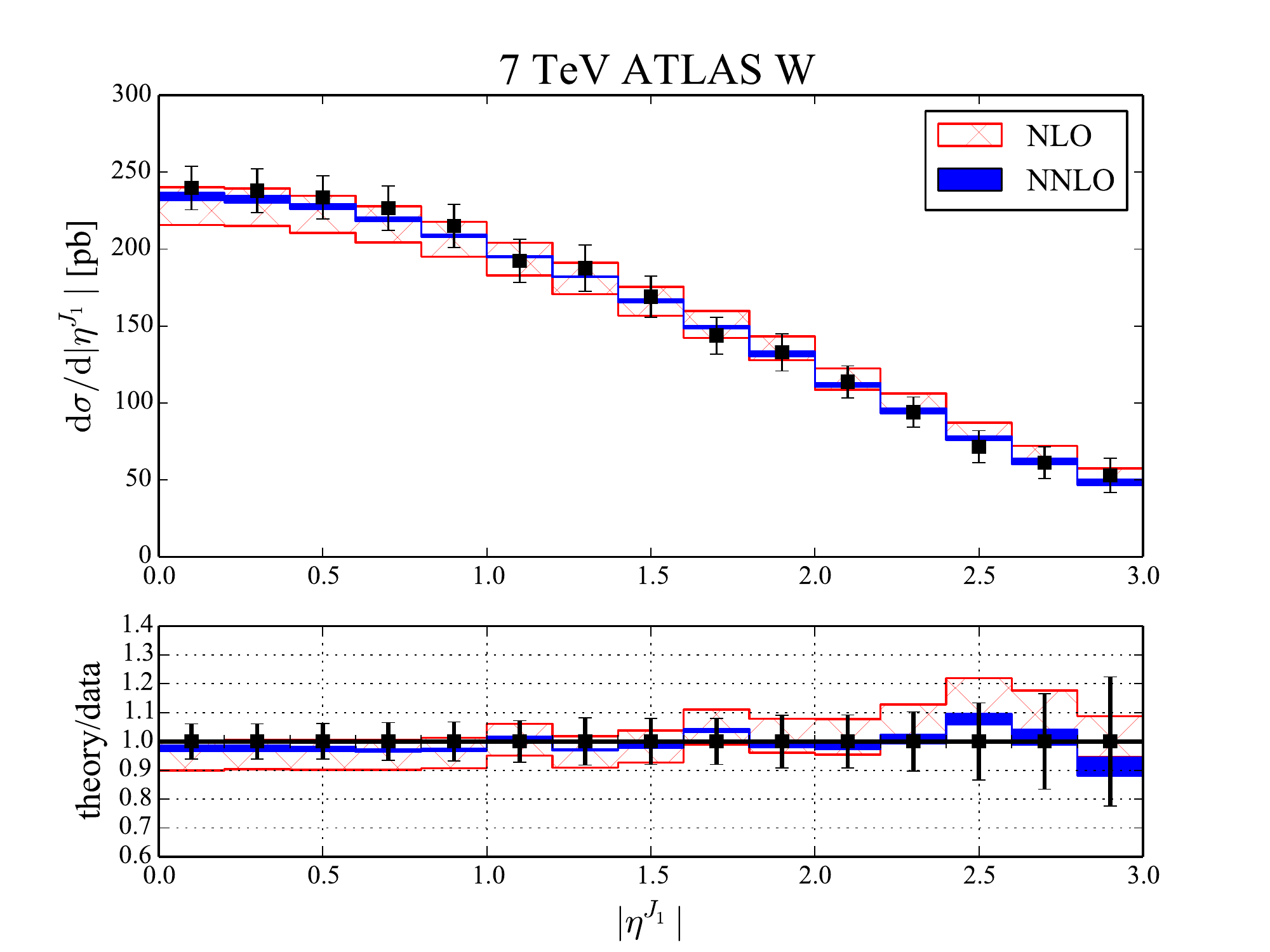} 
    \includegraphics[width=.75\linewidth]{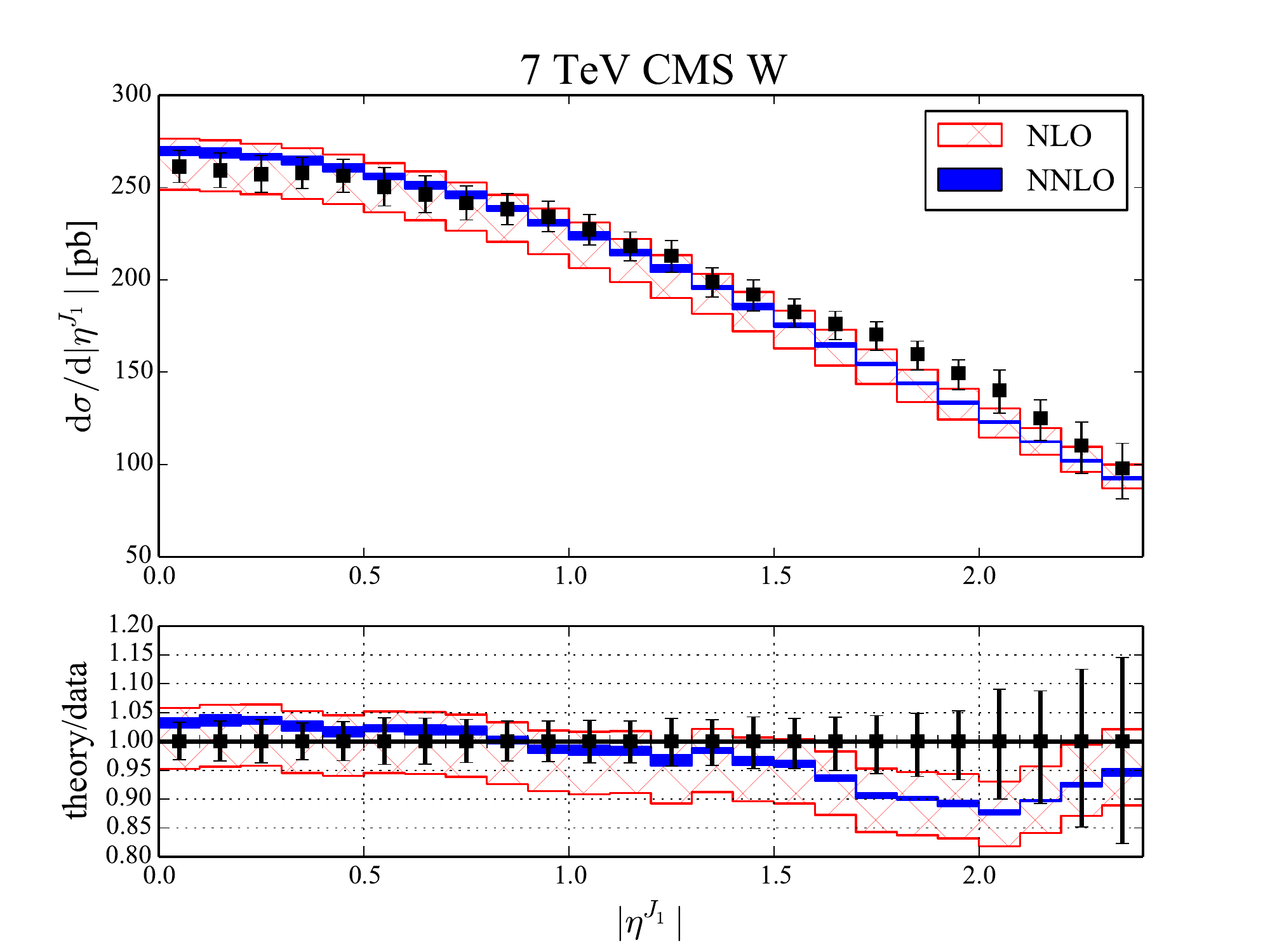} 
    \caption{Plots of the leading-jet rapidity distributions, where both variables are defined as the scalar sum of jet transverse momenta.  The upper panel of each plot shows the distributions at NLO and NNLO, as well as the experimental data points.  The lower panel of each plot shows the ratio of the NLO and NNLO predictions to the measured data.  The black error bars denote the experimental errors, the red hatched band denotes the NLO scale variation, and the blue solid band indicates the NNLO scale variation.}
    \label{fig:etaJ1W}
\end{figure}

Finally, we show in Fig.~\ref{fig:etaJ1W} the comparison between theory and data for the leading-jet rapidity distributions.  The theoretical prediction is in good agreement with the ATLAS data over the entire range.  At small values of rapidity the theoretical prediction is slightly lower than the experimental result, although still within the $1\sigma$ error bars.  At high values of rapidity the NNLO result agrees better with the ATLAS data than the NLO result. ATLAS finds that the {\tt SHERPA}~\cite{Gleisberg:2008ta} merged tree-level prediction tends to overshoot the data at high rapidity, while {\tt ALPGEN}~\cite{Mangano:2002ea} agrees over the the entire range.  The theoretical predictions agree well at central rapidities with the CMS data, but differ slightly at high rapidities.  This discrepancy is seen at both NLO and NNLO, although the size of the difference is only slightly larger than the experimental $1\sigma$ error.  The various tree-level plus parton shower predictions tend to agree slightly better with the data in the high rapidity region.  We have checked that neither the NNPDF 3.0~\cite{Ball:2014uwa} nor the MMHT~\cite{Harland-Lang:2014zoa} parton distribution functions significantly reduce the discrepancy between NNLO QCD and data.  We note that the theory errors are smaller than the experimental errors over the entire rapidity region.

\section{Numerical results for 7 TeV $Z$+jet production}
\label{sec:Znum}

We now discuss $Z$-production in association with a jet, and compare the NLO and NNLO QCD predictions with the results of the ATLAS and CMS collaborations. The fiducial cross sections measured by ATLAS and CMS for the cuts of Table~\ref{tab:zcuts}, as well as the QCD predictions, are shown in Table~\ref{tab:fidZ}.  The NNLO QCD corrections shift the NLO fiducial cross section by $+2\%$ for CMS cuts, and by $-1\%$ for ATLAS cuts.  The residual scale variation is greatly reduced by the inclusion of the NNLO corrections, decreasing from the $\pm 5\%$ level at NLO to below the 1\% level at NNLO.  For both cases the theoretical predictions are in good agreement with the measured fiducial cross sections within the experimental errors.

\begin{table}[h]
\begin{tabular}{|c|c|c||c|}
\hline
$Z$-boson & $\sigma_{\NLO}$ (pb) & $\sigma_{\NNLO}$ (pb) & experiment (pb) \\
 \hline\hline
ATLAS cuts: & $68.9^{+3.4}_{-3.0}$ &  $68.4^{+0.6}_{-1.0}$ & $69.8^{+0.13 \text{(stat)}+ 0.06\,\text{(sys)} +5.22\,\text{(sys)}
	+ 1.26\,\text{(lumi)}}_{-0.13 \text{(stat)}- 0.06\,\text{(sys)} -5.22\,\text{(sys)}
	- 1.26\,\text{(lumi)}}$\\
CMS cuts: & $60.8^{+3}_{-3}$ & $62.0^{+0}_{-0.4}$ & $61.43^{+3.19}_{-3.19}$  \\
\hline
\end{tabular}
\caption{Fiducial cross sections for the inclusive $Z$+1-jet bin for both ATLAS and CMS cuts.  The scale errors are shown for the NLO and NNLO cross sections.}
\label{tab:fidZ}
\end{table} 

\begin{figure}[htbp]
    \includegraphics[width=.75\linewidth]{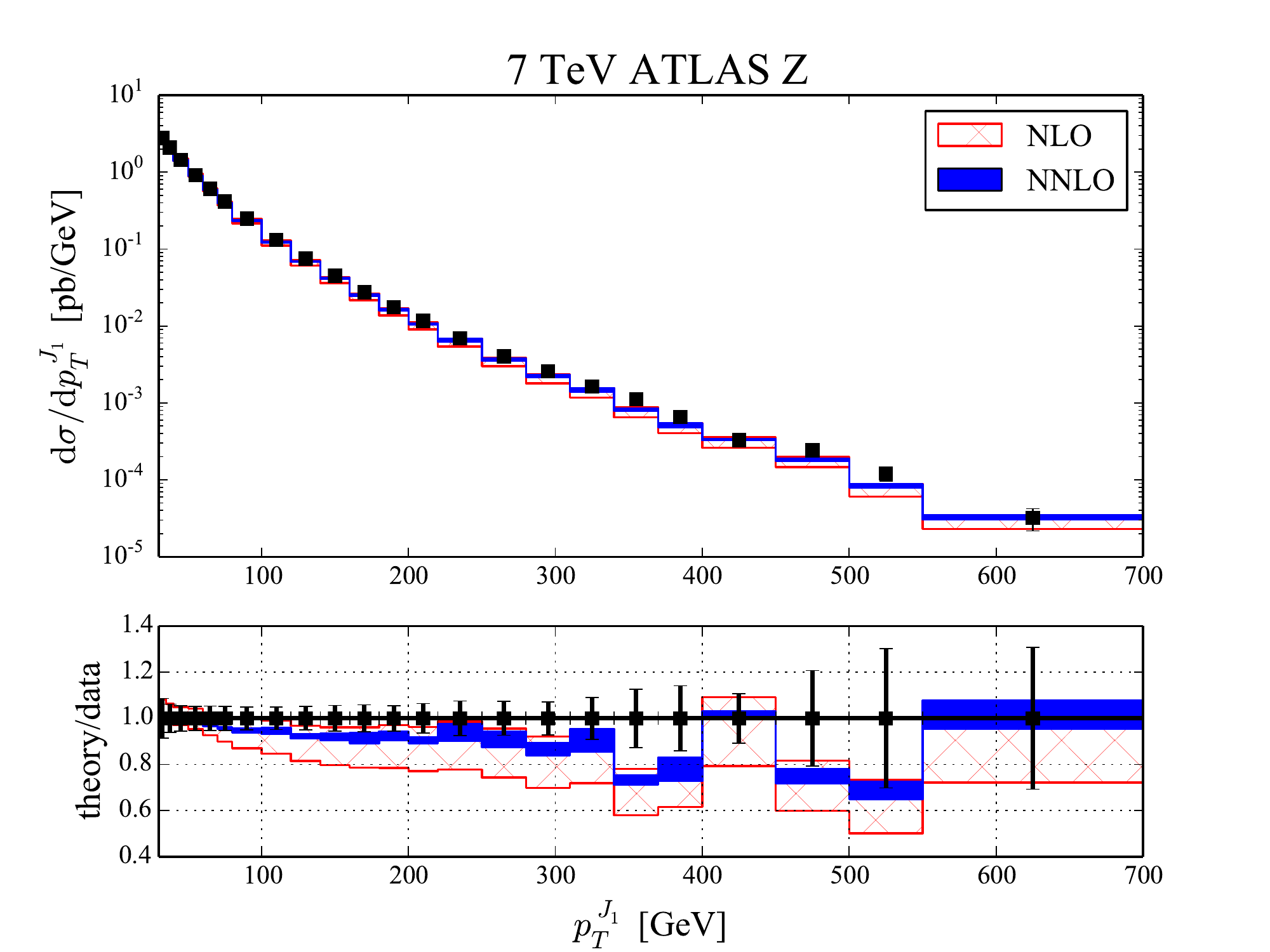} 
    \includegraphics[width=.75\linewidth]{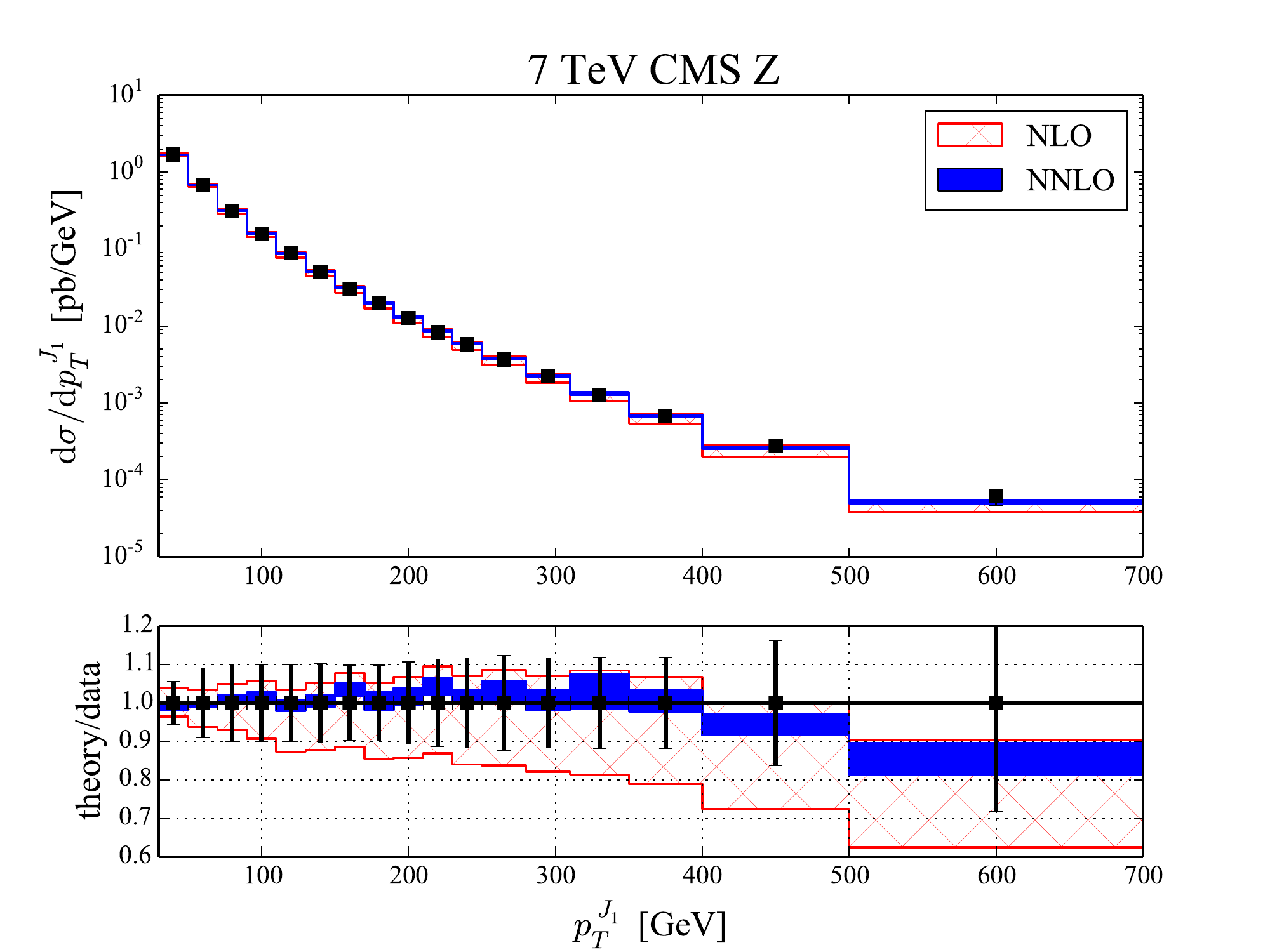} 
    \caption{Plots of the leading-jet transverse momentum distribution for ATLAS and CMS.  The upper panel of each plot shows the distributions at NLO and NNLO, as well as the experimental data points.  The lower panel of each plot shows the ratio of the NLO and NNLO predictions to the measured data.  The black error bars denote the experimental errors, the red hatched band denotes the NLO scale variation, and the blue solid band indicates the NNLO scale variation.}
    \label{fig:pTJ1Z}
\end{figure}

We begin our study of differential distributions in $Z$+jet production with the transverse momentum distribution of the leading jet in Fig.~\ref{fig:pTJ1Z}.  The NNLO QCD prediction is in excellent agreement with the CMS data over the entire $p_T^{J_1}$ range.  The NLO prediction also agrees with the data within errors, but the increase in the cross section at intermediate and high transverse momentum upon inclusion of the NNLO corrections improves the agreement.  The CMS collaboration also compared their measurements against {\tt MADGRAPH}~\cite{Alwall:2011uj} and {\tt POWHEG}~\cite{Frixione:2007vw,Alioli:2010xd} simulations.  Both of these predictions tend to overshoot the data at intermediate and high $p_T^{J_1}$, although {\tt POWHEG} is consistent within the $1\sigma$ experimental errors.  The most satisfactory description of the data is provided by NNLO QCD.

The NNLO QCD prediction is systematically lower than the measured ATLAS data, lying just outside the experimental $1\sigma$ error bars.  At NLO the residual theory error is too large to resolve this slight discrepancy; it becomes apparent only upon the inclusion of the NNLO corrections.  ATLAS also compares their data against tree-level plus parton-shower predictions using {\tt ALPGEN}~\cite{Mangano:2002ea} and {\tt SHERPA}~\cite{Gleisberg:2008ta}.  {\tt ALPGEN} overshoots the data by up to 20\%, while {\tt SHERPA} tends to lie 5-15\% lower.  No prediction gives a completely satisfactory agreement with data. 

\begin{figure}[htbp]
    \includegraphics[width=.75\linewidth]{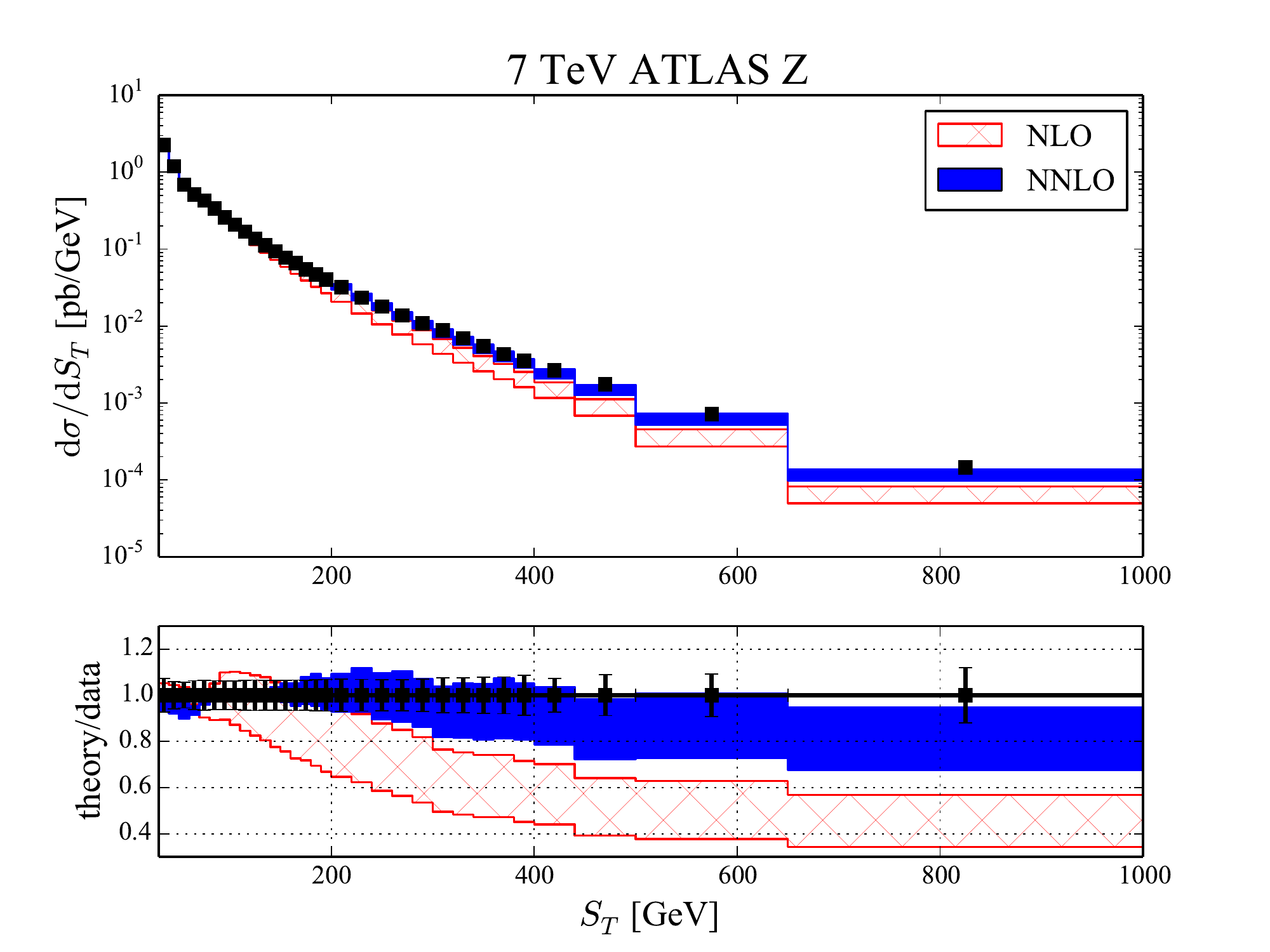} 
    \includegraphics[width=.75\linewidth]{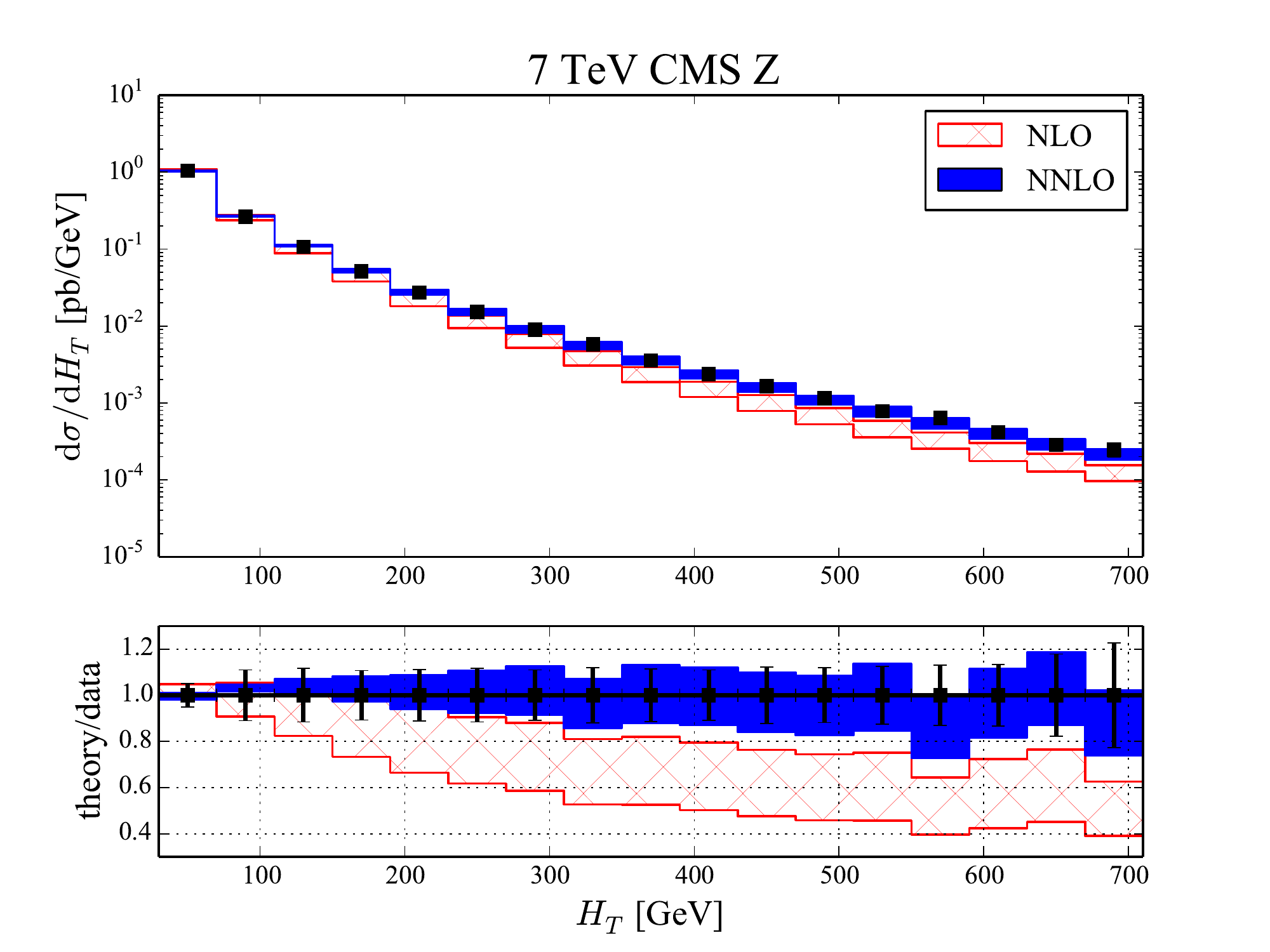} 
    \caption{Plots of the $H_T$ (CMS) and $S_T$ (ATLAS)  distributions, where both variables are defined as the scalar sum of jet transverse momenta.  The upper panel of each plot shows the distributions at NLO and NNLO, as well as the experimental data points.  The lower panel of each plot shows the ratio of the NLO and NNLO predictions to the measured data.  The black error bars denote the experimental errors, the red hatched band denotes the NLO scale variation, and the blue solid band indicates the NNLO scale variation.}
    \label{fig:HTZ}
\end{figure}

The $H_T$ distribution is shown in Fig.~\ref{fig:HTZ}.  For both ATLAS and CMS the NNLO QCD prediction is in excellent agreement with data, with only a small undershoot in the low-$H_T$ region for ATLAS, consistent with the slight difference in the fiducial cross section seen in Table~\ref{tab:fidZ}.  NLO QCD significantly underestimates the cross section at intermediate and high $H_T$, for the same reason as in the $W$+jet case.  Again, to describe the $H_T$ distribution correctly it is essential to have NNLO QCD predictions.  ATLAS compares their data against several tree-level plus parton shower predictions, and finds different results depending on the simulation choice: some give too soft a spectrum, while others predict too hard a spectrum.  An exclusive sum approach pursued by ATLAS gives a different distribution shape than observed in the data.  CMS finds that both {\tt POWHEG} and {\tt MADGRAPH} predict too hard an $H_T$ distribution, by up to 20\% at high $H_T$.  Only NNLO QCD correctly describes the $H_T$ data.  

\begin{figure}[htbp]
    \includegraphics[width=.75\linewidth]{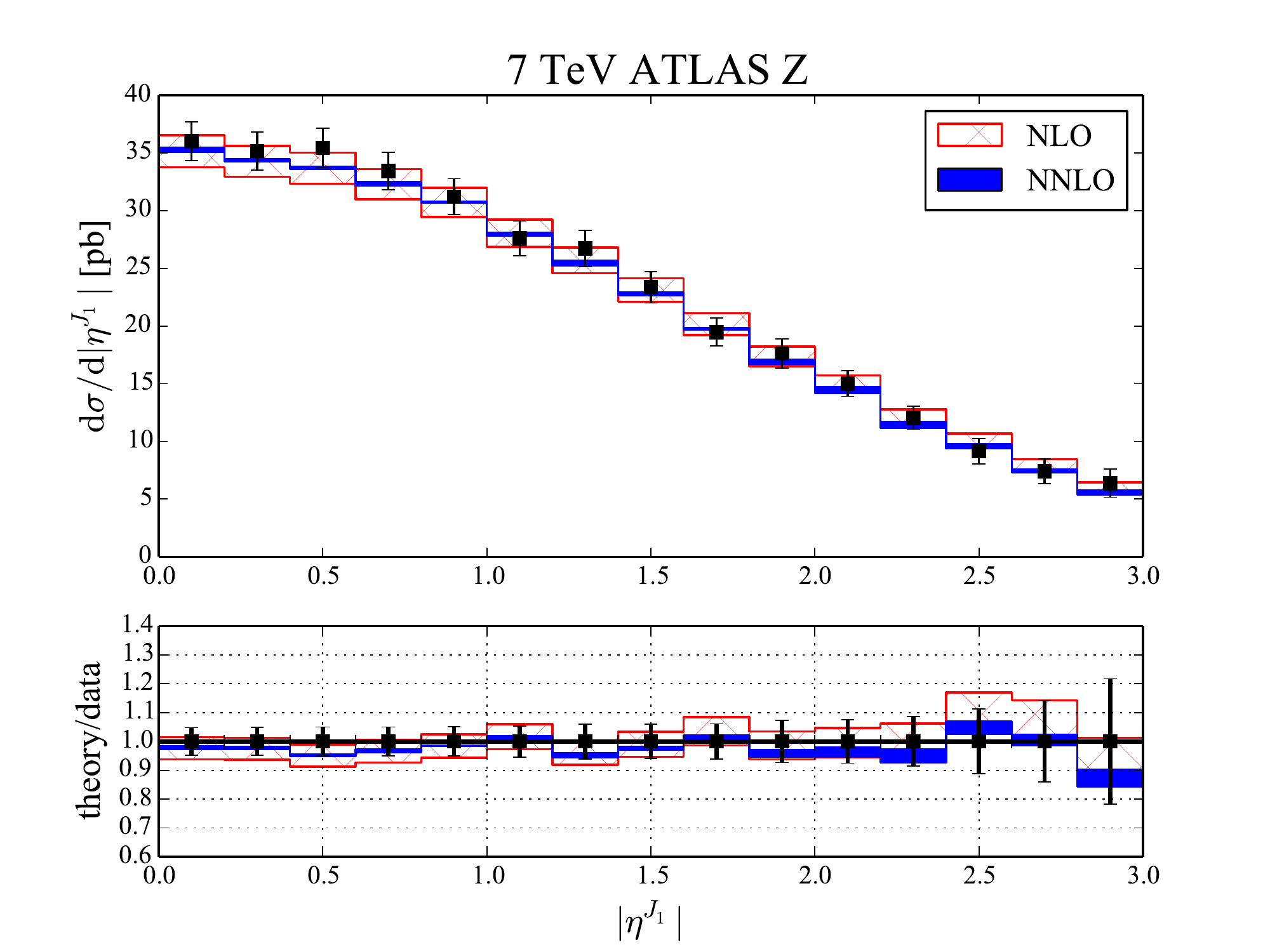} 
    \includegraphics[width=.75\linewidth]{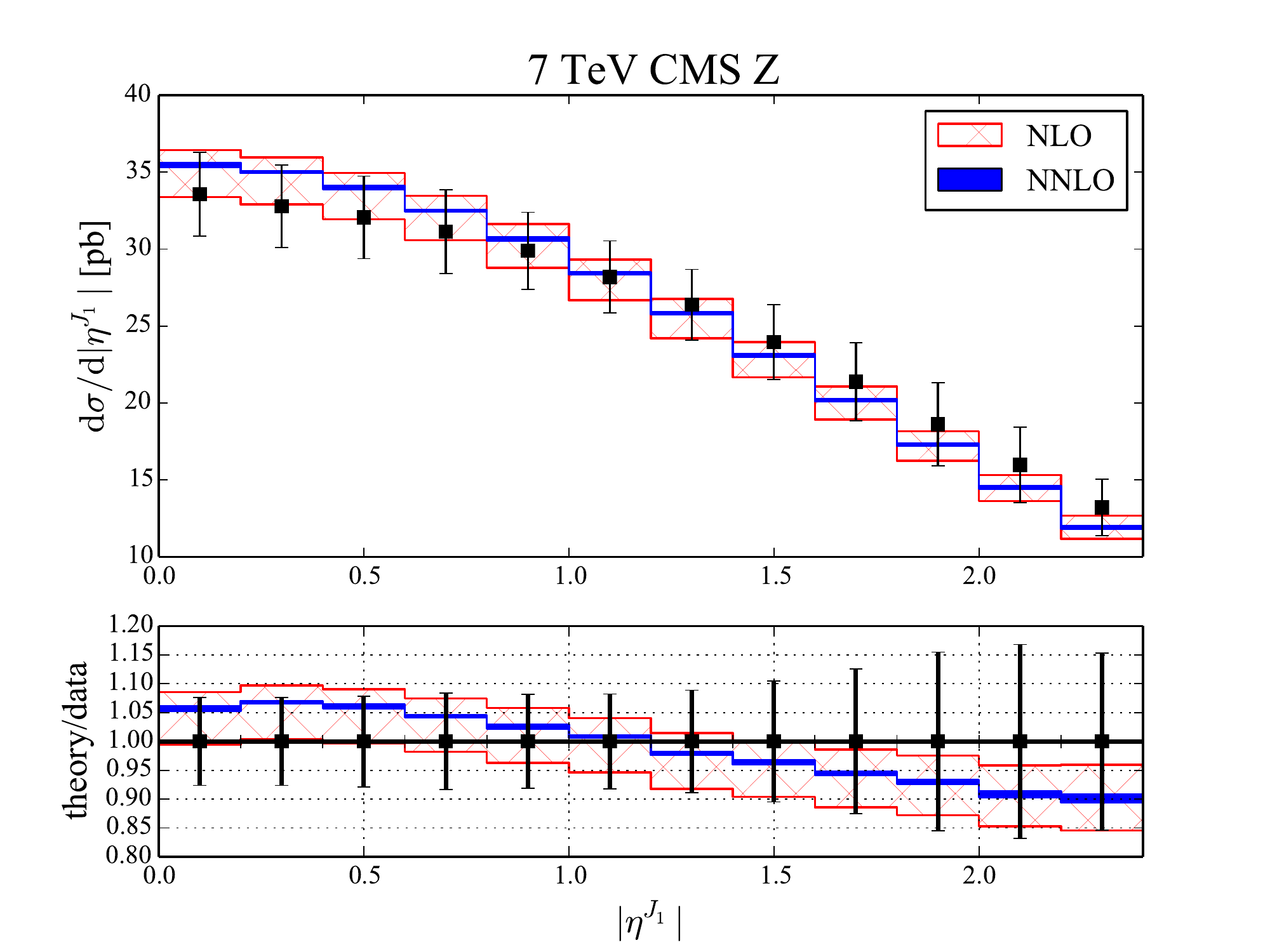} 
    \caption{Plots of the leading-jet rapidity  distributions, where both variables are defined as the scalar sum of jet transverse momenta.  The upper panel of each plot shows the distributions at NLO and NNLO, as well as the experimental data points.  The lower panel of each plot shows the ratio of the NLO and NNLO predictions to the measured data.  The black error bars denote the experimental errors, the red hatched band denotes the NLO scale variation, and the blue solid band indicates the NNLO scale variation.}
    \label{fig:etaJ1Z}
\end{figure}

We next show the rapidity distribution of the leading jet in Fig.~\ref{fig:etaJ1Z}.  NNLO QCD agrees well with the ATLAS data, with only a slight undershoot consistent with the behavior seen for the fiducial cross section.  Although consistent with the $1\sigma$ experimental errors, both NLO and NNLO QCD show a slight shape difference with respect to the CMS data.  Similar small discrepancies are seen by CMS when they compare to {\tt POWHEG} and {\tt MADGRAPH} predictions. 

\begin{figure}[htbp]
    \includegraphics[width=.75\linewidth]{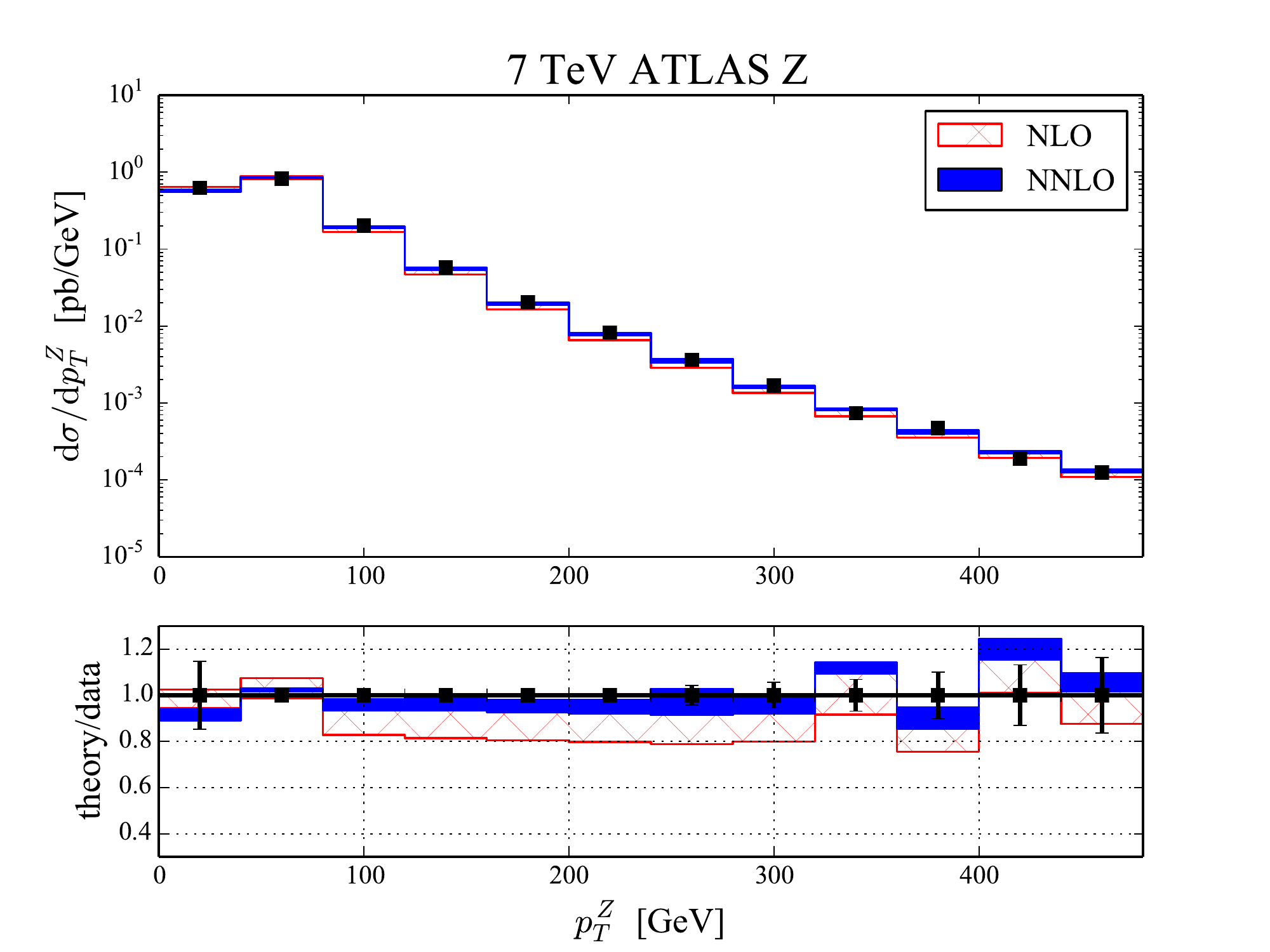} 
    \caption{Plot of the $Z$-boson transverse momentum distribution.  The upper panel shows the distributions at NLO and NNLO, as well as the experimental data points.  The lower panel shows the ratio of the NLO and NNLO predictions to the measured data.  The black error bars denote the experimental errors, the red hatched band denotes the NLO scale variation, and the blue solid band indicates the NNLO scale variation.}
    \label{fig:pTZ}
\end{figure}

Finally, ATLAS has additionally measured the transverse momentum spectrum of the reconstructed $Z$-boson, and we compare QCD predictions to this distribution in Fig.~\ref{fig:pTZ}.  Good agreement between NNLO QCD and the measured distribution is found over the entire range, with only a slight undershoot consistent with the same offset observed for the fiducial cross section.  ATLAS has compared their data against both {\tt ALPGEN} and {\tt SHERPA}.  Both Monte Carlo simulations predict significantly different distribution shapes than seen in the data.  An approach based on exclusive sums used by ATLAS shows a better agreement with the data.

\section{Summary and Conclusions}
\label{sec:conc}

In this paper we have performed a detailed comparison of NNLO QCD predictions with ATLAS and CMS data for the $W$+jet and $Z$+jet processes.  We have studied the fiducial cross sections and numerous distributions: the transverse momentum and rapidity of the leading jet, the jet activity as parameterized by $H_T$, and the transverse momentum distribution of the $Z$-boson. Excellent agreement is observed in almost every observable studied, with only a few small discrepancies lying just outside the experimental $1\sigma$ error bars.  The most notable exception is the intermediate $p_T^{J_1}$ range for $W$ production obtained by CMS, which all available theoretical predictions fail to describe. The most striking observation is the nearly perfect agreement between theory and experiment for the $H_T$ distribution.  This observable has long been theoretically difficult to model.  The search for an accurate modeling of this variable has spurred numerous approximate approaches to higher-order QCD.  For both $W$+jet and $Z$+jet, and for both ATLAS and CMS cuts, NNLO QCD describes this distribution.  In fact, NNLO QCD is the only framework that describes all the available data without significant discrepancies.  

It is worth appreciating that with only simple parton-level predictions one is able to describe all available $V$+jet observables, which span numerous orders of magnitude in both cross section and energy.  It has been previously observed that fixed-order QCD had remarkable power in describing jet cross sections at the Tevatron, despite their apparent complexity and despite all potential complications arising from non-perturbative QCD effects.  This certainly remains the case with the LHC 7 TeV data; for CMS we have applied no correction factors to our fixed-order predictions, while the corrections applied for ATLAS reach a few percent at most in the low transverse momentum region.  After the inclusion of NNLO QCD corrections, which reduce the residual theory uncertainties from uncalculated higher-order QCD corrections to the percent level, the comparison of theory with experiment in the $V$+jet process is limited by the experimental errors.  Future work should combine the electroweak corrections with the NNLO QCD results to facilitate comparisons with Run II data, where electroweak Sudakov effects become increasingly important.  We look forward to even higher precision comparisons upon arrival of the high-luminosity Run II 13 TeV data.

\section*{Acknowledgments}
We thank T.~LeCompte and T.~Childers for helpful discussions.  R.~B. is supported by the DOE contract DE-AC02-06CH11357.  X.~L. is supported by the DOE grant DE-FG02-93ER-40762.  F.~P. is supported by the DOE grants DE-FG02-91ER40684 and DE-AC02-06CH11357.  This research used resources of the National Energy Research Scientific Computing Center, a DOE Office of Science User Facility supported by the Office of Science of the U.S. Department of Energy under Contract No. DE-AC02-05CH11231.  It also used resources of the Argonne Leadership Computing Facility, which is a DOE Office of Science User Facility supported under Contract DE-AC02-06CH11357.

\end{document}